RU98-8-B
%
%
\headline{\hfil \folio}
\hoffset=0.5truein
\hsize=5.5truein
\vsize=8truein
%
\catcode`@=11                           
\newskip\ttglue
\def\ninefonts{%
   \global\font\ninerm=cmr9%
   \global\font\ninei=cmmi9%
   \global\font\ninesy=cmsy9%
   \global\font\nineex=cmex10%
   \global\font\ninebf=cmbx9%
   \global\font\ninesl=cmsl9%
   \global\font\ninett=cmtt9%
   \global\font\nineit=cmti9%
   \skewchar\ninei='177%
   \skewchar\ninesy='60%
   \hyphenchar\ninett=-1%
   \moreninefonts
   \gdef\ninefonts{\relax}}%
\def\moreninefonts{\relax}                      


\def\elevenfonts{%
   \global\font\elevenrm=cmr10 scaled \magstephalf%
   \global\font\eleveni=cmmi10 scaled \magstephalf%
   \global\font\elevensy=cmsy10 scaled \magstephalf%
   \global\font\elevenex=cmex10%
   \global\font\elevenbf=cmbx10 scaled \magstephalf%
   \global\font\elevensl=cmsl10 scaled \magstephalf%
   \global\font\eleventt=cmtt10 scaled \magstephalf%
   \global\font\elevenit=cmti10 scaled \magstephalf%
   \global\font\elevenss=cmss10 scaled \magstephalf%
   \skewchar\eleveni='177%
   \skewchar\elevensy='60%
   \hyphenchar\eleventt=-1%
   \moreelevenfonts
   \gdef\elevenfonts{\relax}}%
\def\moreelevenfonts{\relax}

\def\twelvefonts{%
   \global\font\twelverm=cmr10 scaled \magstep1%
   \global\font\twelvei=cmmi10 scaled \magstep1%
   \global\font\twelvesy=cmsy10 scaled \magstep1%
   \global\font\twelveex=cmex10 scaled \magstep1%
   \global\font\twelvebf=cmbx10 scaled \magstep1%
   \global\font\twelvesl=cmsl10 scaled \magstep1%
   \global\font\twelvett=cmtt10 scaled \magstep1%
   \global\font\twelveit=cmti10 scaled \magstep1%
   \global\font\twelvess=cmss10 scaled \magstep1%
   \skewchar\twelvei='177%
   \skewchar\twelvesy='60%
   \hyphenchar\twelvett=-1%
   \moretwelvefonts
   \gdef\twelvefonts{\relax}}%
\def\moretwelvefonts{\relax}                    

\def\fourteenfonts{%
   \global\font\fourteenrm=cmr10 scaled \magstep2%
   \global\font\fourteeni=cmmi10 scaled \magstep2%
   \global\font\fourteensy=cmsy10 scaled \magstep2%
   \global\font\fourteenex=cmex10 scaled \magstep2%
   \global\font\fourteenbf=cmbx10 scaled \magstep2%
   \global\font\fourteensl=cmsl10 scaled \magstep2%
   \global\font\fourteenit=cmti10 scaled \magstep2%
   \global\font\fourteenss=cmss10 scaled \magstep2%
   \skewchar\fourteeni='177%
   \skewchar\fourteensy='60%
   \morefourteenfonts
   \gdef\fourteenfonts{\relax}}%
\def\morefourteenfonts{\relax}                  


\def\tenmibfonts{
   \global\font\tenmib=cmmib10%
   \global\font\tenbsy=cmbsy10%
   \skewchar\tenmib='177%
   \skewchar\tenbsy='60%
   \gdef\tenmibfonts{\relax}}

\def\elevenmibfonts{
   \global\font\elevenmib=cmmib10 scaled \magstephalf%
   \global\font\elevenbsy=cmbsy10 scaled \magstephalf%
   \skewchar\elevenmib='177%
   \skewchar\elevenbsy='60%
   \gdef\elevenmibfonts{\relax}}

\def\twelvemibfonts{
   \global\font\twelvemib=cmmib10 scaled \magstep1%
   \global\font\twelvebsy=cmbsy10 scaled \magstep1%
   \skewchar\twelvemib='177%
   \skewchar\twelvebsy='60%
   \gdef\twelvemibfonts{\relax}}

\def\fourteenmibfonts{
   \global\font\fourteenmib=cmmib10 scaled \magstep2%
   \global\font\fourteenbsy=cmbsy10 scaled \magstep2%
   \skewchar\fourteenmib='177%
   \skewchar\fourteenbsy='60%
   \gdef\fourteenmibfonts{\relax}}

\def\mib{
   \tenmibfonts%
   \textfont0=\tenbf\scriptfont0=\sevenbf%
   \scriptscriptfont0=\fivebf%
   \textfont1=\tenmib\scriptfont1=\seveni%
   \scriptscriptfont1=\fivei%
   \textfont2=\tenbsy\scriptfont2=\sevensy%
   \scriptscriptfont2=\fivesy}%

\def\ninepoint{\ninefonts               
   \def\rm{\fam0\ninerm}%
   \textfont0=\ninerm\scriptfont0=\sevenrm\scriptscriptfont0=\fiverm
   \textfont1=\ninei\scriptfont1=\seveni\scriptscriptfont1=\fivei
   \textfont2=\ninesy\scriptfont2=\sevensy\scriptscriptfont2=\fivesy
   \textfont3=\nineex\scriptfont3=\nineex\scriptscriptfont3=\nineex
   \textfont\itfam=\nineit\def\it{\fam\itfam\nineit}%
   \textfont\slfam=\ninesl\def\sl{\fam\slfam\ninesl}%
   \textfont\ttfam=\ninett\def\tt{\fam\ttfam\ninett}%
   \textfont\bffam=\ninebf
   \scriptfont\bffam=\sevenbf
   \scriptscriptfont\bffam=\fivebf\def\bf{\fam\bffam\ninebf}%
   \def\mib{\relax}%
   \tt\ttglue=.5emplus.25emminus.15em
   \normalbaselineskip=11pt
   \setbox\strutbox=\hbox{\vrule height 8pt depth 3pt width 0pt}%
   \normalbaselines\rm\singlespaced}%

\def\tenpoint{
   \def\rm{\fam0\tenrm}%
   \textfont0=\tenrm\scriptfont0=\sevenrm\scriptscriptfont0=\fiverm
   \textfont1=\teni\scriptfont1=\seveni\scriptscriptfont1=\fivei
   \textfont2=\tensy\scriptfont2=\sevensy\scriptscriptfont2=\fivesy
   \textfont3=\tenex\scriptfont3=\tenex\scriptscriptfont3=\tenex
   \textfont\itfam=\tenit\def\it{\fam\itfam\tenit}%
   \textfont\slfam=\tensl\def\sl{\fam\slfam\tensl}%
   \textfont\ttfam=\tentt\def\tt{\fam\ttfam\tentt}%
   \textfont\bffam=\tenbf
   \scriptfont\bffam=\sevenbf
   \scriptscriptfont\bffam=\fivebf\def\bf{\fam\bffam\tenbf}%
   \def\mib{%
      \tenmibfonts%
      \textfont0=\tenbf\scriptfont0=\sevenbf%
      \scriptscriptfont0=\fivebf%
      \textfont1=\tenmib\scriptfont1=\seveni%
      \scriptscriptfont1=\fivei%
      \textfont2=\tenbsy\scriptfont2=\sevensy%
      \scriptscriptfont2=\fivesy}%
   \tt\ttglue=.5emplus.25emminus.15em
   \normalbaselineskip=12pt
   \setbox\strutbox=\hbox{\vrule height 8.5pt depth 3.5pt width 0pt}%
   \normalbaselines\rm\singlespaced}%

\def\elevenpoint{\elevenfonts           
   \def\rm{\fam0\elevenrm}%
   \textfont0=\elevenrm\scriptfont0=\sevenrm\scriptscriptfont0=\fiverm
   \textfont1=\eleveni\scriptfont1=\seveni\scriptscriptfont1=\fivei
   \textfont2=\elevensy\scriptfont2=\sevensy\scriptscriptfont2=\fivesy
   \textfont3=\elevenex\scriptfont3=\elevenex\scriptscriptfont3=\elevenex
   \textfont\itfam=\elevenit\def\it{\fam\itfam\elevenit}%
   \textfont\slfam=\elevensl\def\sl{\fam\slfam\elevensl}%
   \textfont\ttfam=\eleventt\def\tt{\fam\ttfam\eleventt}%
   \textfont\bffam=\elevenbf
   \scriptfont\bffam=\sevenbf
   \scriptscriptfont\bffam=\fivebf\def\bf{\fam\bffam\elevenbf}%
   \def\mib{%
      \elevenmibfonts%
      \textfont0=\elevenbf\scriptfont0=\sevenbf%
      \scriptscriptfont0=\fivebf%
      \textfont1=\elevenmib\scriptfont1=\seveni%
      \scriptscriptfont1=\fivei%
      \textfont2=\elevenbsy\scriptfont2=\sevensy%
      \scriptscriptfont2=\fivesy}%
   \tt\ttglue=.5emplus.25emminus.15em
   \normalbaselineskip=13pt
   \setbox\strutbox=\hbox{\vrule height 9pt depth 4pt width 0pt}%
   \normalbaselines\rm\singlespaced}%

\def\twelvepoint{\twelvefonts\ninefonts 
   \def\rm{\fam0\twelverm}%
   \textfont0=\twelverm\scriptfont0=\ninerm\scriptscriptfont0=\sevenrm
   \textfont1=\twelvei\scriptfont1=\ninei\scriptscriptfont1=\seveni
   \textfont2=\twelvesy\scriptfont2=\ninesy\scriptscriptfont2=\sevensy
   \textfont3=\twelveex\scriptfont3=\twelveex\scriptscriptfont3=\twelveex
   \textfont\itfam=\twelveit\def\it{\fam\itfam\twelveit}%
   \textfont\slfam=\twelvesl\def\sl{\fam\slfam\twelvesl}%
   \textfont\ttfam=\twelvett\def\tt{\fam\ttfam\twelvett}%
   \textfont\bffam=\twelvebf
   \scriptfont\bffam=\ninebf
   \scriptscriptfont\bffam=\sevenbf\def\bf{\fam\bffam\twelvebf}%
   \def\mib{%
      \twelvemibfonts\tenmibfonts%
      \textfont0=\twelvebf\scriptfont0=\ninebf%
      \scriptscriptfont0=\sevenbf%
      \textfont1=\twelvemib\scriptfont1=\ninei%
      \scriptscriptfont1=\seveni%
      \textfont2=\twelvebsy\scriptfont2=\ninesy%
      \scriptscriptfont2=\sevensy}%
   \tt\ttglue=.5emplus.25emminus.15em
   \normalbaselineskip=14pt
   \setbox\strutbox=\hbox{\vrule height 10pt depth 4pt width 0pt}%
   \normalbaselines\rm\singlespaced}%

\def\fourteenpoint{\fourteenfonts\twelvefonts 
   \def\rm{\fam0\fourteenrm}%
   \textfont0=\fourteenrm\scriptfont0=\twelverm\scriptscriptfont0=\tenrm
   \textfont1=\fourteeni\scriptfont1=\twelvei\scriptscriptfont1=\teni
   \textfont2=\fourteensy\scriptfont2=\twelvesy\scriptscriptfont2=\tensy
   \textfont3=\fourteenex\scriptfont3=\fourteenex
      \scriptscriptfont3=\fourteenex
   \textfont\itfam=\fourteenit\def\it{\fam\itfam\fourteenit}%
   \textfont\slfam=\fourteensl\def\sl{\fam\slfam\fourteensl}%
   \textfont\bffam=\fourteenbf
   \scriptfont\bffam=\twelvebf
   \scriptscriptfont\bffam=\tenbf\def\bf{\fam\bffam\fourteenbf}%
   \def\mib{%
      \fourteenmibfonts\twelvemibfonts\tenmibfonts%
      \textfont0=\fourteenbf\scriptfont0=\twelvebf%
      \scriptscriptfont0=\tenbf%
      \textfont1=\fourteenmib\scriptfont1=\twelvemib%
      \scriptscriptfont1=\tenmib%
      \textfont2=\fourteenbsy\scriptfont2=\tenbsy%
      \scriptscriptfont2=\tenbsy}%
   \normalbaselineskip=17pt
   \setbox\strutbox=\hbox{\vrule height 12pt depth 5pt width 0pt}%
   \normalbaselines\rm\singlespaced}%
%
%

\def\singlespaced{
   \baselineskip=\normalbaselineskip}           


%
%
\twelvepoint
%
%

\def\showheadline#1#2{\headline={\ifnum\pageno>1{\ifodd\pageno{\hfil\tenpoint #1\hfil} %
\else{\hfil\tenpoint #2\hfil}\fi} \else{\hfil}\fi}}

\def\address#1{\hbox to \hsize{\hglue 0.29in\relax
\vbox{\hsize=4.70in\relax\rightskip=0pt plus 1in\relax\noindent#1}\hfil}}

\long\def\beginaddress#1\endaddress{\vglue 6pt\address{#1}\vglue 24pt}

\def\finalversion{\headline{\hfil}}

\def\section#1{\vskip 24pt plus4pt minus4pt\goodbreak\leftline{\bf #1}%
\vglue 12pt\nobreak\noindent\kern -0.0em}

\def\subsection#1{\vskip 12pt plus4pt minus4pt\goodbreak\leftline{\bf #1}%
\nobreak\noindent\kern -0.0em}

\def\subsubsection#1{\vskip 12pt plus4pt minus4pt\goodbreak\leftline{\it #1}%
\nobreak\noindent\kern -0.0em}

\def\begincaption#1{\begingroup\tenpoint\noindent#1\ \ \ }
\def\endcaption{\endgroup}

\newbox\@capbox                                 
\newcount\@caplines                             

\def\references{\section{REFERENCES}\tenpoint\parindent=0pt
\raggedright\rightskip=0pt plus 5em}

\def\ref#1#2{\hbox to \hsize{\vbox{\tenpoint\hsize=0.2in\relax #1\hfil}
\hfil\vtop{\hsize=4.75in\relax\tenpoint #2}}}
%
%
\vglue 1.0truein

\finalversion
%
%
\input epsf
%
%
%
%
\bigskip
\centerline{\bf{Boson-Fermion Model of High-$T_C$ Superconductivity}}
\centerline{{\bf{----a Progress Report}}\footnote*{An invited lecture at 
XXII International School of Theoretical Physics, Ustron, Poland, 
Sept. 10-15, 1998}}
\bigskip
\centerline{\it Hai-cang Ren}
\bigskip
\centerline{National Center for Theoretical Science,} 
\centerline{National Tsing Hua University, Hsinchu, Taiwan, ROC}
\centerline{and}
\centerline{Department of Physics, The Rockefeller University,}
\centerline{New York, NY 10021, USA\footnote\dag{permenant address}}
\bigskip
\bigskip
\bigskip
\section{1. The Motivation}

A common feature of the cuprate superconductors is that the coherence 
length is comparable with the lattice spacing [1]. 
Therefore the Cooper pairs are rather localized in the 
coordinate space and consequently can be regarded as boson degrees of 
freedom [2]. The superconductivity is thereby associated with the 
kinematical Bose-Einstein condensation and 
there must be uncondensed pairs in the normal phase. 

For a Bose-Einstein condensation to happen, the thermal wavelength of the 
bosons with mass $m_b$ at the transition temperature $T_C$, $\lambda_C=
\sqrt{{2\pi \over m_b k_BT_C}}$ should be comparable 
with the inter-boson distance, $l$, so that quantum mechanical coherence 
takes place. Indeed, the ratio ${\lambda_C\over l}$ is 1.38 for an ideal Bose 
gas and 1.65 for $HeII$ at the $\lambda$-transition. For high $T_C$ materials, 
this ratio can be extracted from the results of the $\mu SR$ experiment [3] of 
Uemura et. al. According to them, the transition temperature $T_C$ of the 
under-doped up to optimal-doped cuprates is inversely 
proportional to the square of the magnetic penetration depth at $T=0$, 
$\lambda_0$ with a constant of proportionality universal for all cuprates, 
$$T_C({\rm K})=0.25\times10^5{m_e\over e^2\lambda_0^2},\eqno(1.2)$$ 
where $m_e$ is the electron mass in vacuum and the unit of 
$m_e/(e^2\lambda_0^2)$ is $\AA^{-3}$. Taking into account of possible 
contribution to the superfluid density from the fermionic component, we have
$${1\over \lambda_0^2}\geq {4e^2\over m_bcl^2},\eqno(1.3)$$ where $c$ denotes
the average distance between $CuO_2$ layers and $l$ stands for the average 
inter-boson distance within each layer. Combining (1.3) and (1.2), we 
end up with $${\lambda_C\over l}\leq 2.8,\eqno(1.4)$$ which is consistent 
with the picture of Bose-Einstein condensation.

With simply an ideal electron(hole) gas in chemical equilibrium with an 
ideal gas of Cooper pairs, we are able to explain Uemura's universal 
dependence of $T_C$ on the superfluid density [4].

\section{2. The Boson-Fermion Model}

The boson-fermion model was proposed independently in Ref. 5 and in 
Ref. 3. The grand Hamiltonian of the system reads 
$$H=\sum_{\vec p,s}\epsilon_{\vec p}a_{\vec ps}^\dagger a_{\vec ps}
+\sum_{\vec p}\omega_{\vec p}b_{\vec p}^\dagger b_{\vec p}$$
$$+{1\over \sqrt{\Omega}}\sum_{\vec p,\vec q}g_{\vec p,\vec q}
(b_{\vec p}a_{{\vec p\over 2}+\vec q\uparrow}^\dagger
a_{{\vec p\over 2}-\vec q\downarrow}^\dagger+
b_{\vec p}^\dagger a_{{\vec p\over 2}-\vec q\downarrow}
a_{{\vec p\over 2}+\vec q\uparrow}),\eqno(2.1)$$ where
$a_{\vec ps}$, $a_{\vec ps}^\dagger$, $b_{\vec p}$ and 
$b_{\vec p}^\dagger$ are annihilation and creation operators of 
electrons(holes) and bosons, the subscript $s$ denotes the spin orientation,
and $\Omega$ is the total volume of the system. 
The momentum dependence of the coupling $g_{\vec p,\vec q}$ will determine 
the pairing symmetry, e. g., $g={\rm{const}}\to s$-wave and $g\propto 
q_x^2-q_y^2\to d$-wave. To illustrate the main physics, we assume a $3D$ 
isotropic jellium in which
$$\epsilon_{\vec p}={p^2\over 2m_f}-\mu\eqno(2.2)$$ and
$$\omega_{\vec p}={p^2\over 2m_b}+2(\nu-\mu)\eqno(2.3)$$ and a constant $g$, 
where $m_f(m_b)$ is the effective mass of fermions(bosons),
$\mu$ is the chemical potential of the system, $2\nu$ is the energy of 
a static boson relative to two static fermions 
($2\nu=\omega_{\vec p}|_{\vec p=0}-2\epsilon_{\vec p}|_{\vec p=0}$), 

The conserved electric charge number is given by
$$Q=\sum_{\vec p,s}a_{\vec ps}^\dagger a_{\vec ps}
+2\sum_{\vec p}b_{\vec p}^\dagger b_{\vec p}.\eqno(2.4)$$
Because of the strong Coulomb repulsion, we assume that 
$\nu>0$ and the bosons are resonances. A dimensionless coupling
can be defined as the ratio of the boson half width  
in vacuum, $\Gamma/2$, to the energy $2\nu$, i.e.
$$\hat g^2={\Gamma\over 2\nu}={g^2\over\pi}
\Big({m_f\over 2}\Big)^{3\over 2}{1\over\sqrt{\nu}},\eqno(2.5)$$ 
which serves as an effective expansion parameter. 

The physics of the model at weak coupling $\hat g<<1$ and $T=0$ 
is determined by the total number density of charges,
$$n=n_f+2n_b\eqno(2.6)$$ in relation to the characteristic density
$$n_\nu={(2m_f\nu)^{3\over 2}\over 3\pi^2},\eqno(2.7)$$ where $n_f$ 
is the number density of fermions and $n_b$ that of bosons. 
$n_\nu$ corresponds to a filled Fermi sea with Fermi energy $\nu$.
If $n<n_\nu$, the electrons(holes) will fill in the Fermi levels below 
$\nu$ and there can only be virtual bosons through interaction condensating 
at the zero momentum bosonic level. The long range order in this case
is of BCS type. If $n>n_\nu$, the excess electrons(holes) 
after filling up the Fermi level $\nu$ prefer to combine into bosons 
which condensing at the zero momentum level and the long range 
order is of Bose-Einstein type. 

The grand partition function of the model (2.1) is 
$${\cal Q}={\rm{Tr}}e^{-\beta H}\eqno(2.8)$$ and the thermal average of an 
operator $O$ is defined to be
$$<O>={{\rm{Tr}}Oe^{-\beta H}\over {\rm{Tr}}e^{-\beta H}}.\eqno(2.9)$$ 
With a Bose condensate
$$B={1\over\sqrt{\Omega}}<b_{\vec p=0}>,\eqno(2.10)$$ 
we may substitute 
$$b_{\vec p}=\sqrt{\Omega}B\delta_{\vec p,0}+\beta_{\vec p}\eqno(2.11)$$
into the Hamiltonian (2.1) and obtain $H=H_0+H_1$, with 
$$H_0=2\Omega(\nu-\mu)B^2+\sum_{\vec p,s}\epsilon_{\vec p}a_{\vec ps}^\dagger 
a_{\vec ps}+\sum_{\vec p}\omega_{\vec p}\beta_{\vec p}^\dagger\beta_{\vec p}$$
$$+gB\sum_{\vec q}(a_{\vec q\uparrow}^\dagger
a_{-\vec q\downarrow}^\dagger+a_{-\vec q\downarrow}
a_{\vec q\uparrow})\eqno(2.12)$$ and $H_1$ the rest of the terms of (2.1) which 
are of higher orders at weak coupling. $B$ has been chosen real. 
The Hamiltonian (2.12) can be easily diagonalized through a Bogoliubov 
transformation and yields a fermionic spectrum
$$E_{\vec p}=\sqrt{\epsilon_{\vec p}^2+\Delta^2}\eqno(2.13)$$ 
with a gap energy $\Delta=gB$. The thermodynamic potential within this 
approximation reads
$$\ln{\cal Q}=2\beta(\mu-\nu)B^2\Omega+2\sum_{\vec p}
\ln(1+e^{-\beta E_{\vec p}})$$ $$-\sum_{\vec p}\ln(1-e^{-\beta(\omega_{\vec p}
+2\nu-2\mu)}).\eqno(2.14)$$ The thermodynamical equilibrium corresponds to 
the maximum of $\ln{\cal Q}$ at fixed $\beta$ and $\mu$, i.e.
$$\Big({\partial\ln{\cal Q}\over\partial B^2}\Big)_{\beta,\mu}=0\eqno(2.15)$$
and $$\Big({\partial^2\ln{\cal Q}\over(\partial B^2)^2}\Big)_{\beta,\mu}
\leq 0.\eqno(2.16)$$ Combining (2.15) and the relation
$$n={1\over\Omega}\Big({\partial\ln{\cal Q}\over\partial\mu}\Big)_{\beta,B}
,\eqno(2.17)$$ we can solve for the gap energy $\Delta(T)$ and the transition 
temperature $T_C$ in terms of the density $n$. 

At low density, $n<n_\nu$, we find that 
$$\Delta(0)=8\mu\exp\Big(-2-{\nu-\mu\over\hat g^2\sqrt{\nu\mu}}\Big)
\eqno(2.18)$$ with $\mu\simeq{(3\pi^2n)^{2\over 3}\over 2m}$ and that
$${\Delta(0)\over k_BT_C}=\pi e^\gamma\eqno(2.19)$$ with 
$\gamma=0.5772...$ the Eular constant. At higher density, $n>n_\nu$, 
we obtain that $$k_BT_C={2\pi m_b}\Big({n-n_\nu\over 2\zeta({3\over 2})
}\Big)^{2\over 3}\eqno(2.20)$$ with $\zeta(3/2)=2.612...$ and that
$$\Delta(T)=\Delta(0)\sqrt{1-\Big({T\over T_C}\Big)^{3\over 2}}\eqno(2.21)$$ 
with $\Delta(0)=g\sqrt{(n-n_\nu)/2}$. Therefore, the phenomenological 
model (2.1) provides a simple interpolation between a BCS 
condensation and a Bose-Einstein condensation.

\section{3. The Pseudo-gap}

A remarkable phenomenon of under-doped cuprates is the pseudo gap in 
their electron(hole) spectrum above $T_C$ [6]. It has been found 
numerically in the boson-fermion model [7]. Here I shall present  
an analytical calculation [8], which highlight the importance of the 
dimensionality of the system. 

Starting with the retarded electron propagator
$$S_R(p_0,\vec p)=-i\int_0^\infty dte^{ip_0t}<\{a_{\vec p}(t),
a_{\vec p}^\dagger(0)\}>,\eqno(3.1)$$ where the time 
development of $a_{\vec p}(t)$ and its conjugate follows from the total 
Haniltonian (2.1), the spectral function $A(p_0,\vec p)$ is given by 
its imaginary, i.e. 
$$A(p_0,\vec p)={1\over\pi}{\rm{Im}}S_R(p_0,\vec p),\eqno(3.2)$$ The function 
$A(p_0,\vec p)$ gives rise to the probability density of a single electron 
(hole) excitation at energy $p_0$ and momentum $\vec p$ and can be measured 
directly by ARPES. For a BCS superconductor, $A(p_0,\vec p)$ is peaked at the 
quasi particle pole $p_0=\epsilon_{\vec p}$ in the normal phase and at 
$p_0=\pm\sqrt{\epsilon_{\vec p}^2+\Delta^2}$ below $T_C$. 

1). Two-dimensions: This case simulates the actual cuprate materials 
above $T_C$, which consist of a set of parallel $CuO_2$ layers. 
As long as $T$ is not too close to $T_C$, the interlayer hopping can be 
neglected. With a jellium approximation, $\epsilon_{\vec p}$ and 
$\omega_{\vec p}$ remain given by (2.2) and (2.3) but with a $2D$ momentum 
$\vec p$. On writing $$S_R(p_0,\vec p)={1\over p_0-\epsilon_{\vec p}
-\Sigma_R(p_0,\vec p)},\eqno(3.3)$$ the retarded self-energy function, 
$\Sigma_R(p_0,\vec p)$, is given to the one-loop order by
$$\Sigma_R(p_0,\vec p)=g^2\int{d^2q\over (2\pi)^2}{N_b(\vec q)
+N_f(\vec q-\vec p)
\over p_0-\omega_{\vec q}+\epsilon_{\vec q-\vec p}+i0^+}\eqno(3.4)$$
with $$N_b(\vec p)={1\over e^{\beta\omega_{\vec p}}-1}\eqno(3.5)$$ and
$$N_f(\vec p)={1\over e^{\beta\epsilon_{\vec p}}+1}\eqno(3.6)$$ 
respectively. There is no Bose-Einstein condensation in a truly $2D$
system, the bosonic chemical potential $2(\mu-\nu)\equiv-\delta$ vanishes 
at $T=0$ and the integral (3.4) diverges logarithmically. For an 
approximate $2D$ system and with a fixed carrier density,
$\delta<<k_BT$ within a considerable range of $T<<\nu/k_B$ above $T_C$ [8], 
and the integral (3.4) is dominated at small
$\vec q$. This gives rise to a crude estimate that 
$$\Sigma_R(p_0,\vec p)\sim {\bar\Delta^2\over p_0+\epsilon_{\vec p}}
\eqno(3.7)$$ with
$$\bar\Delta^2={2r\over \pi}\hat g^2\nu k_B T\ln{k_B T\over\delta}.
\eqno(3.8)$$ and $r=m_b/m_f$, which leads to a perfect gap. 
A more refined calculation of the integration (3.4) was performed in [8] 
for $p_0\sim k_BT$ with the result at $p=p_F$ (Fermi momentum):
$$A(p_0,\vec p)|_{p=p_F}={1\over\pi}{\rm{Im}}{1\over p_0-u(p_0)-iv(p_0)
+u(\delta)},\eqno(3.9)$$ where
$$u(p_0)={2r\over \pi}\hat g^2\nu\Bigg[{k_BT{\rm sign}(p_0-\delta)\over
\sqrt{4r\mu\delta+(p_0-\delta)^2}}\ln{\sqrt{4r\mu\delta
+(p_0-\delta)^2}+|p_0-\delta|\over \sqrt{4r\mu\delta+(p_0-\delta)^2}-
|p_0-\delta|}$$ $$-{\ln r\over r-1}
+{1\over 2}f(e^{-\beta(p_0-\delta)})\sqrt{{\pi k_BT\over 
r\mu}}\Bigg]\eqno(3.10)$$ and $$v(p_0)=-{2r\over \pi}\hat g^2\nu\Bigg[
{\pi k_BT\over\sqrt{4r\mu\delta+(p_0-\delta)^2}}
+{1\over 2}\Big(f(e^{\beta(p_0-\delta)})+\zeta\Big({1\over 2}\Big)\Big)
\sqrt{{\pi k_BT\over r\mu}}\Bigg]\eqno(3.11)$$ with 
$$f(z)={2\over\sqrt{\pi}}\int_0^\infty dx\sqrt{x}{ze^{-x}\over (1+ze^{-x})^2}
\eqno(3.12)$$ and $\zeta(1/2)=-1.4604...$. The function (3.9) is plotted in 
Fig. 1. 

\topinsert
\hbox to\hsize{\hss
	\epsfxsize=4.0truein\epsffile{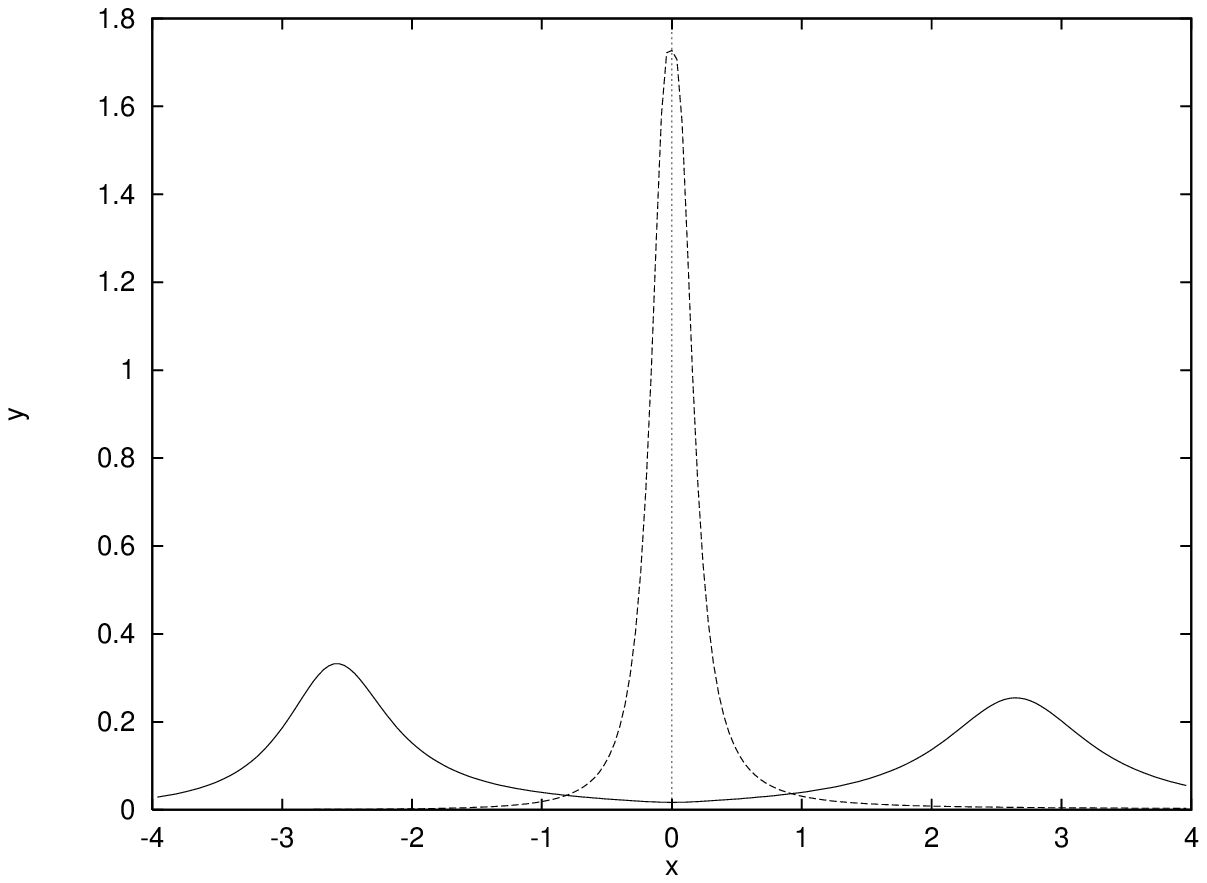}
	\hss}
\begincaption{Figure 1}
The spectral function at $p=p_F$, with $\beta\mu=10$, 
$r=2$, $\ln k_B T/\delta=8$ and $\hat g^2={\pi\over 80}$ for $D=2$ 
(solid line) and $D=3$ (dashed line), where
$x=p_0/k_BT$ and $y=A(p_0,\vec p)|_{p=p_F}k_BT$.
\endcaption
\endinsert

The argument based on the logarithmic divergence can be generalized to 
higher orders in $g$ and a Borel summation of the most divergent diagrams 
yields [8]: $$A(p_0,\vec p)|_{p=p_F}={|p_0|\over\bar\Delta^2}e^{-{p_0^2
\over\bar\Delta^2}}.\eqno(3.13)$$ It consists of two peaks at $p_0=\pm
\bar\Delta/\sqrt{2}$ and a depletion of states at the Fermi level $p_0=0$.

2). Three dimensions: With three dimensional $\vec p$ and $\vec q$, the simple 
argument based on the logarithmic divergence does not work since the 
integral (3.4) converges as $\delta\to 0$. The evaluation of the integral 
for $3D$ case leads to 
$$\Sigma(p_0,\vec p)|_{p=p_F}={\pi r\over 4}\hat g^2k_BT
+{r\over\pi}\hat g^2\mu \int_0^{\infty}dx{x\over e^{\beta\mu(x^2-1)}+1}$$
$$-{ir\over 4}\hat g^2k_BT\Big[\ln{k_BT\over\delta+
{(p_0-\delta)^2 \over 4\mu}}+2\ln(1+e^{\beta(p_0-\delta)})\Big].\eqno(3.14)$$
The corresponding $A(p_0,\vec p)$ function is also plotted in Fig. 1 for 
comparison and the pseudo-gap disappears. Since the $3D$ 
calculation is only to the one-loop order, this is not in contradiction 
with a weak pseudo-gap, as inferred from the numerical solutions of truncated 
Dyson-Schwinger equations [9]. 

The dependence of the pseudo-gap can be tested by changing the separation 
between $CuO_2$ layers, or by comparison with $3D$ strongly correlated 
superconductors, say fullerenes perhaps. 

\section{4. DC Transport Coefficients}

On the normal phase DC transport coefficients (resistivity, Hall number and 
magnetoresistance), lies perhaps, the most severe disagreement between 
the observation and the conventional Fermi liquid theory. A large number 
of measurements converges to a universal temperature dependence above 
$T_C$ for all cuprates [10]. The resistivity $\rho$ depends linearly on $T$ 
in contrast with the $T^2$ behavior of a good 
metal. The Hall number $n_H$, instead of being a constant which equals to 
the actual carrier density in case of a good metal, increases linearly 
with $T$ and reaches a value about a factor two of the chemically determined 
carrier density. The magnetoresistance, $\Delta\rho/\rho$ being proportional 
to $T^{-n}$ ($n=3.5-4$)is at entirely variance with the Kholer's law
(which says that the ratio ${\delta\rho\over\rho}/{B^2\over\rho^2}$ is 
temperature independent) for good metals. 

In the boson-fermion model, there are two kinds of charge carriers 
above $T_C$: electrons or holes (fermions) and uncondensed Cooper pairs
(bosons). If the lifetime of the latter is longer than the relaxation 
time of the electric current, both of them contribute to the DC conductivity
A Green's function calculation is, however, very difficult since 
the simplifications based on incoherent superposition of 
quasi elastic collisions fail for bosons. Instead, we ask the following 
question [11]: {\it{Given a chemical equilibrium of fermions and bosons 
($g=0$ in (2.1)) and a simple power dependence on $T$ for the relaxation 
time of each kind of carriers, Is it possible to produce a reasonable fit 
of available experimental data?}}

Consider the situation with an electric field $\vec E$ parallel to the 
$CuO_2$ plane and a magnetic field $\vec B=B\hat z$ perpendicular to it, 
the electric current density is given by [11]
$$J_a=\sigma_{ab}(\vec B)E_b\eqno(4.1)$$ with $a,b=x,y$. The Taylor 
expansion in $B$ of the conductivity tensor reads
$$\sigma_{ab}(\vec B)=(\sigma+\Delta\sigma)\delta_{ab}+\sigma_H\epsilon_{ab}
+O(B^3),\eqno(4.2)$$ where 
$$\sigma=\Big({\tau_fn_f\over m_f}+4{\tau_bn_b\over m_b}\Big)e^2,\eqno(4.3)$$ 
is the zero field conductivity, 
$$\sigma_H=\Big(\eta{\tau_f^2n_f\over m_f^2}+8{\tau_b^2n_b\over m_b^2}\Big)e^3B
\eqno(4.4)$$ is the Hall conductivity and $$\Delta\sigma=\Big(\eta^\prime
{\tau_f^3n_f\over m_f^3}+8{\tau_b^3n_b\over m_b^3}
\Big)e^4B^2,\eqno(4.5)$$ with $n_f$ and $n_b$ the number densities of each 
type of carriers, $\tau_f$ and $\tau_b$ the corresponding relaxation times, 
and $m_f$ and $m_b$ the corresponding effective masses. We have assumed a 
parabolic band for bosons and a nonparabolic band for fermions. In case of 
the nonparabolic band, the effective mass $m_f$ is defined by 
$${1\over m_f}={1\over 2n_f}\int{d^2\vec k\over (2\pi)^2}v_{\vec k}^2
\delta(\epsilon_{\vec k})\eqno(4.6)$$ and the band correction 
factors $\eta$ and $\eta^\prime$ of (4.4) and (4.5) are given by 
$$\eta={m_f^2\over 2n_f}\epsilon_{ab}\epsilon_{ij}
\int{d^2\vec k\over (2\pi)^2}v_av_i\nabla_jv_b\delta
(\epsilon_{\vec k})\eqno(4.7)$$ and 
$$\eta^\prime={m_f^3\over 2n_f}\epsilon_{ab}\epsilon_{ij}
\int{d^2\vec k\over (2\pi)^2}v_a\nabla_bv_lv_i\nabla_jv_l\delta
(\epsilon_{\vec k}).\eqno(4.8)$$ It follows from (4.3)-(4.5) that 
the DC resistivity 
$$\rho={1\over e^2}\Big({\tau_fn_f\over m_f}+4{\tau_bn_b\over m_b}\Big)^{-1}
,\eqno(4.9)$$ the Hall number $$n_H={\sigma^2\over e\sigma_H}B=
{\Big({\tau_fn_f\over m_f}+4{\tau_bn_b\over m_b}\Big)^2\over
\eta{\tau_f^2n_f\over m_f^2}+8{\tau_b^2n_b\over m_b^2}}\eqno(4.10)$$ and the 
Kholer's ratio $$K=-\rho^3{\Delta\sigma+\rho\sigma_H^2\over B^2}=
{e^4\tau_fn_f\rho^3\over m_f}\Big[(\eta^\prime-\eta^2){\tau_f^2\over m_f^2}
+{4e^2\tau_bn_b\over m_b}\Big(\eta{\tau_f\over m_f}-{2\tau_b\over m_b}\Big)^2
\rho\Big].\eqno(4.11)$$ 

The approximate agreement with Luttinger theorem as observed by photo 
emission ruled out any issue based on a large fraction of bosons and we 
expect that $n_b<<n_f$. The temperature dependence of $n_f$ and $n_b$ 
can be neglected. The DC resistivity is dominated by fermions and 
we have $$\tau_f\propto T^{-1}.\eqno(4.12)$$ An explanation of this non
-Fermi liquid behavior is suggested in [12]. If the factor $\eta$ is 
sufficiently small such that the denominator of (4.10) is dominated by the 
second term, a Hall number considerably larger than the actual carrier 
density emerges and the linear temperature dependence can be obtained if 
we assume $$\tau_b\propto T^{-1.5}.\eqno(4.13)$$ With both (4.12) and (4.13), 
the Kholer's ratio would be proportional to $T^{-1.5}$ if the factor 
$\eta^\prime$ is also sufficiently small. For a nonparabolic band, it is 
possible to have small $\eta$ since it switches sign from a particle 
Fermi sea to a hole Fermi sea. But a small $\eta^\prime$ may not be easy to 
tune since its integrand is positive definite. We are still not in the 
position to claim the triumph of the boson-fermion model in explaining the 
DC transport coefficients.

\noindent
\section{5. Two Theoretical Issues}

To justify the boson-fermion model as an adequate low density 
phenomenological model of certain strongly correlated electronic systems, 
the following issues have to be settled.

\noindent
{\it{5.1 Fermi distribution function}}

Consider a purely electronic system on a lattice with electronic operators 
$\alpha_{\vec p,s}$ and $\alpha_{\vec p,s}^\dagger$ in the momentum
representation. The Fermi distribution function is defined as 
$$n_{\vec p}=\sum_s<|\alpha_{\vec p,s}^\dagger\alpha_{\vec p,s}|>
\eqno(5.1)$$ with $|>$ the ground state of the system. 
With the Bogoliubov type of trial state [13]
$$|>=\exp\Big[\sum_{\vec p}\theta_{\vec p}(\alpha_{-\vec p,\downarrow}
\alpha_{\vec p,\uparrow}-\alpha_{\vec p,\uparrow}^\dagger
\alpha_{-\vec p,\downarrow}^\dagger)\Big]|0>,\eqno(5.2)$$ we find that
$$n_{\vec p}=2\sin^2\theta_{\vec p},\eqno(5.3)$$ 

For a BCS ground state, we have
$$\cos2\theta_{\vec p}={\epsilon_{\vec p}\over \sqrt{\epsilon_{\vec p}^2
+\Delta^2}}\eqno(5.4)$$ and $$\sin2\theta_{\vec p}={\Delta\over 
\sqrt{\epsilon_{\vec p}^2+\Delta^2}}\eqno(5.5)$$ with $\Delta<<\epsilon_F$
and the corresponding $n_{\vec p}$ is slightly smeared by $\Delta$ from 
that of an ideal Fermi gas. The location of the kink (Fermi surface 
when $\Delta=0$) is determined by Luttinger theorem.

For a Hubbard model with a strong on-site attraction, the ground state 
is the Bose condensate of local pairs. This corresponds to a $\vec p$ 
independent $\theta_{\vec p}$ and the corresponding Fermi distribution 
function is a plateau within the Brillouin zone.

In case of the boson-fermion model, the distribution function  
corresponds to an different interpolation from that in [13]
between the former two cases. The $n_{\vec p}$ profile 
is the superposition of the BCS case with the Fermi level {\it{retreated}} 
towards the bottom of the band and with a plateau outside the Fermi sea. 

The Fermi distribution function for the three different cases is 
shown schematically in Fig. 2. The profile in Fig. 2c  
may be generated with a Hubbard like model with an on-site 
attraction and a nearest neighbor repulsion [14].

\topinsert
\hbox to\hsize{\hss
	\epsfxsize=4.0truein\epsffile{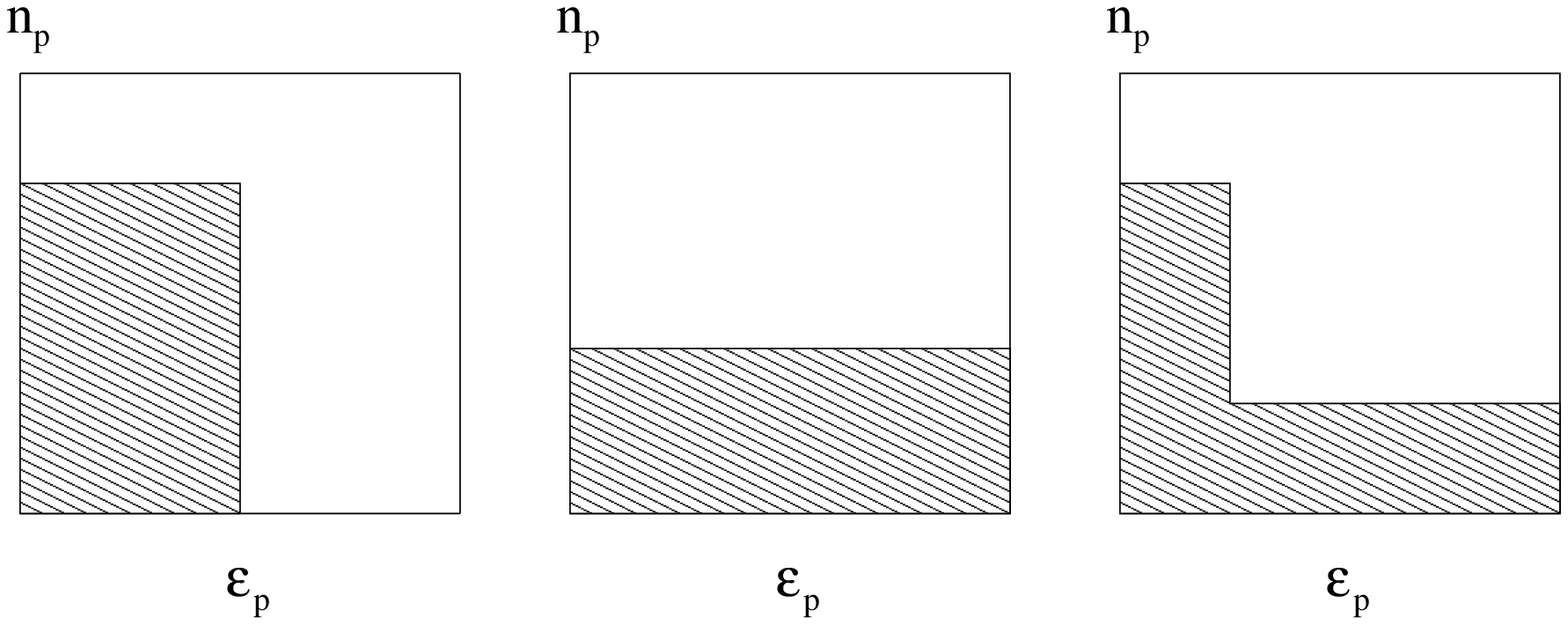}
	\hss}
\begincaption{Figure 2}
The schematic Fermi distribution function at $T=0$ for BCS model (a), 
local pair model (b) and boson-fermion model (c), where the gap smearing 
effect is not shown. 
\endcaption
\endinsert

\noindent
{\it{5.2 The pole trajectory underneath the physical sheet at $T>T_C$}}

In the BCS model, the pairing instability would be triggered at the absence 
of a long range order by a pair of complex conjugate poles of the two 
fermion scattering amplitude 
below $T_C$ on the energy $E$-plane. For $T>T_C$, they slip through the cut 
at $E=2\epsilon_F$ to the unphysical sheets below and leave the real axis 
{\it{vertically}} with an increasing $T$. Their effect becomes insignificant 
outside the critical region. On the other hand, for the local pair system  
the bound state pole stays always on the real $E$ axis for 
$T>T_C$. The pole trajectory for 
the boson-fermion model (2.1) (for which the pole of the  
scattering amplitude of two fermions coincides with the pole of 
the boson propagator ) has been calculated 
to the one-loop order. Without a Bose condensate, there would be a pair of 
complex conjugate poles triggering the instability for $T<T_C$. As $T$ being 
increased away from $T=T_C$, they emigrant to the unphysical sheet but 
leave the real axis {\it{obliquely}} with a {\it{tunable}} slope proportional 
to $g^2$ (Fig. 3). For small $g$, they stay close to the real $E$ for a wide 
range of temperature above $T_C$ and could make significant contributions
to transport processes. Starting with a {\it{purely electronic}} Hamiltonian, 
such resonance boson poles can be implemented for a few body system [14]. 
But to reproduce the pole trajectory of the boson-fermion model 
for a many-body system remains an open problem.

\topinsert
\hbox to\hsize{\hss
	\epsfxsize=4.0truein\epsffile{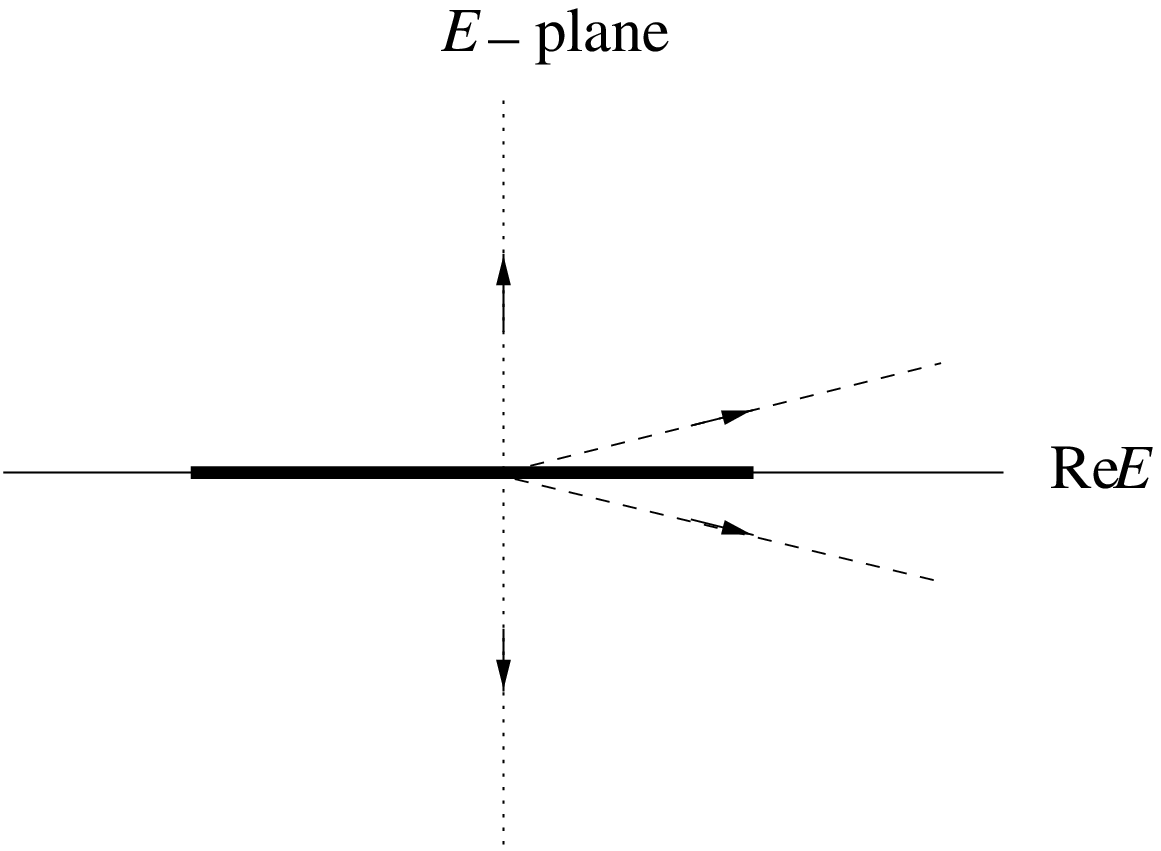}
	\hss}
\begincaption{Figure 3}
The pole trajectory on the complex energy plane of the two electron
(hole) scattering amplitude for $T>T_C$. The dashed line is for the 
boson-fermion model and the dotted line is for BCS model. The arrows
point to the direction corresponding to an increasing $T$ The heavy 
solid line represents the branch cut and the intersection point of all 
trajectories corresponds to $E=2\epsilon_F$. 
\endcaption
\endinsert
\bigskip

I am grateful to Professor R. Micnas and the organizing committee
for inviting me to this prestigious conference. I am also benefited 
from many discussions with Professors R. Micnas and 
S. Robaszkiewicz. This work is supported in part by National Science 
Council of ROC under Grant NSC-CTS-981003 and by U. S. Department of 
Energy under Grant DE-FG02-91ER40651, Task B.

\references

\ref{1.}{P. Chaudhari, {\it et. al.}, {\it Phys. Rev.}, {\bf B36}, 8903 
(1987).}
\ref{2.}{Y. Uemura, {\it et. al.}, {\it Phys. Rev. Lett. } {\bf 62}, 2317 
(1989); Y. Uemura, "Energy Scales of High $T_C$ Cuprates, Doped-Fullerenes, 
and Other Exotic Superconductors", in {\it "High-$T_C$ Superconductivity and 
$C_{60}$ Family"}, Proceeding of CCAST Symposium/Workshop, ed. S. Q. Feng 
and H. C. Ren, Gordon and Breach Pub. Inc., 1994.}
\ref{3.}{R. Friedberg and T. D. Lee, {\it Phys. Lett.} {\bf A138}, 423 
(1989); {\it Phys. Rev. B} {\bf 40}, 6745 (1989).}
\ref{4.}{R. Friedberg, T. D. Lee and H. C. Ren, {\it Phys. Lett. A} 
{\bf 152}, 423 (1991)}
\ref{5.}{J. Ranninger and S. Robaszkiewicz, {\it Physica} {\bf B135}, 
468 (1985); See also the review article by R. Micnas, J. Ranninger and 
S. Robaszkiewicz, {\it Rev. Mod. Phys.} {\bf 62}, 113 (1990), and the 
references therein.}
\ref{6.}{A. G. Loeser, D. S. Dessau and Z. X. Shen, {\it Physica} 
{\bf C263}, 208 (1996); H. Ding, {\it et. al.}, 
{\it Nature}, {\bf Vol. 382}, 51 (1996); A. G. Loeser, {\it et. al.} 
{\it Science}, {\bf Vol. 273}, 325 (1996).}
\ref{7.}{J. Ranninger, J. M. Robin and M. Eschrig, {\it Phys. Rev. Lett.}, 
{\bf 74}, 4027 (1995).}
\ref{8.}{Hai-cang Ren, {\it Physica} {\bf C303}, 115 (1998).}
\ref{9.}{J. Maly, B. Janko and K. Levin, cond-mat/9805018.} 
\ref{10.}{N. P. Ong, Y. F. Yan and J. M. Harris, "Charge Transport Properties 
of Cuprate Superconductors", in {\it "High-$T_C$ Superconductivity and 
$C_{60}$ Family}, Proceeding of CCAST Symposium/Workshop, ed. S. Q. Feng 
and H. C. Ren, Gordon and Breach Pub. Inc., 1994.} 
\ref{11.}{Hai-cang Ren, to be published.}
\ref{12.}{J. M. Robin, A. Romano and J. Ranninger, {\it{Phys. Rev. Lett.}}
{\bf 81}, 2755 (1998).}
\ref{13.}{P. Nozieres and S. Schmitt-Rink, {\it J. Low. Temp. Phys.}, 
{\bf 59}, 195 (1985).}
\ref{14.}{R. Friedberg, T. D. Lee and H. C. Ren, {\it Phys. Rev. B} {\bf 50}, 
10190 (1994).}

\vfill\eject
\end

\bye